\documentclass[sigconf]{acmart}

\usepackage{booktabs} % For formal tables
\usepackage{amssymb}
\usepackage{graphicx}
\usepackage{etoolbox}
\usepackage{caption}
\usepackage{amsmath}
\usepackage{multirow}
\usepackage{color}
\usepackage{url}
\usepackage{enumerate}
\usepackage{balance}
\usepackage{array}
\usepackage{appendix}
\usepackage{subfig}
\usepackage{svg}

\copyrightyear{2018} 
\acmYear{2018} 
\setcopyright{acmcopyright}
\acmConference[WebSci'18]{10th ACM Conference on Web Science}{May 27--30, 2018}{Amsterdam, Netherlands}
\acmBooktitle{WebSci'18: 10th ACM Conference on Web Science, May 27--30, 2018, Amsterdam, Netherlands}
\acmPrice{15.00}
\acmDOI{10.1145/3201064.3201067}
\acmISBN{978-1-4503-5563-6/18/05}

\begin{document}
\title{Wisdom in Sum of Parts: Multi-Platform Activity \\ Prediction in Social Collaborative Sites}

\author{Roy Ka-Wei Lee}
\affiliation{
  \institution{Living Analytics Research Centre\\ Singapore Management University }
}
\email{roylee.2013@smu.edu.sg}

\author{David Lo}
\affiliation{
	\institution{Singapore Management University }
}
\email{davidlo@smu.edu.sg}

\begin{abstract}
In this paper, we proposed a novel framework which uses user interests inferred from activities (a.k.a.,\textit{ activity interests}) in multiple social collaborative platforms to predict users' platform activities. Included in the framework are two prediction approaches: (i) \textit{direct platform activity prediction}, which predicts a user's activities in a platform using his or her activity interests from the same platform (e.g., predict if a user answers a given Stack Overflow question using the user's interests inferred from his or her prior \textit{answer} and \textit{favorite} activities in Stack Overflow), and (ii) \textit{cross-platform activity prediction}, which predicts a user's activities in a platform using his or her activity interests from another platform (e.g., predict if a user answers a given Stack Overflow question using the user's interests inferred from his or her \textit{fork} and \textit{watch} activities in GitHub). To evaluate our proposed method, we conduct prediction experiments on two widely used social collaborative platforms in the software development community: GitHub and Stack Overflow. Our experiments show that combining both \textit{direct} and \textit{cross} platform activity prediction approaches yield the best accuracies for predicting user activities in GitHub (AUC=0.75) and Stack Overflow (AUC=0.89). 
\end{abstract}


\maketitle

\section{Introduction}
\label{sec:introduction}
Software developers are increasingly adopting social collaborative platforms for software development. \textit{GitHub} and \textit{Stack Overflow} are two of such popular platforms. GitHub is a collaborative software development platform that allows code sharing and version control. Users can participate in various activities in GitHub, for example, users may \textit{fork} (i.e., create a copy of) repositories of other users or \textit{watch} the activities of repositories of interest. Stack Overflow is a technical question-and-answer community-based website where users post and answer questions relating to software development. 

As these social collaborative platforms gain popularity, many research studies have proposed recommender systems to improve the usability of these platforms. For example, there are work which predict and recommend relevant Stack Overflow questions and answers to aid users in software development \cite{deSouza2014,Wang2015,wang2016}. While for GitHub, researchers have proposed methods to predict which software repositories are more relevant to a target user \cite{guendouz2015,zhang2014,jiang2017}. Nevertheless, many of these studies only consider the users' behaviours and interests in a single platform when predicting and recommending user platform activities.

There have been few existing inter-platform studies on GitHub and Stack Overflow. Vasilescu et al. \cite{vasilescu2013} studied how users' involvement in Stack Overflow impacted their productivity in GitHub. Badashian et al. \cite{badashian2014} did an empirical study on the correlation between different types of user activities in GitHub and Stack Overflow. In a more recent study by Lee and Lo \cite{lee2017}, the researchers found that users who have accounts on both GitHub and Stack Overflow do share similar interests across the two platforms. For example, a user who commits to Java-related repositories in GitHub, is likely to also answer Java-related questions in Stack Overflow. In this paper, we aim to extend the study in \cite{lee2017}, and propose a multi-platform activity prediction method, which predicts a user's activities in a platform using the user's interests inferred from his or her activities in multiple platforms. 

\begin{figure*}[h]
	\centering
	\includegraphics[scale = 0.5]{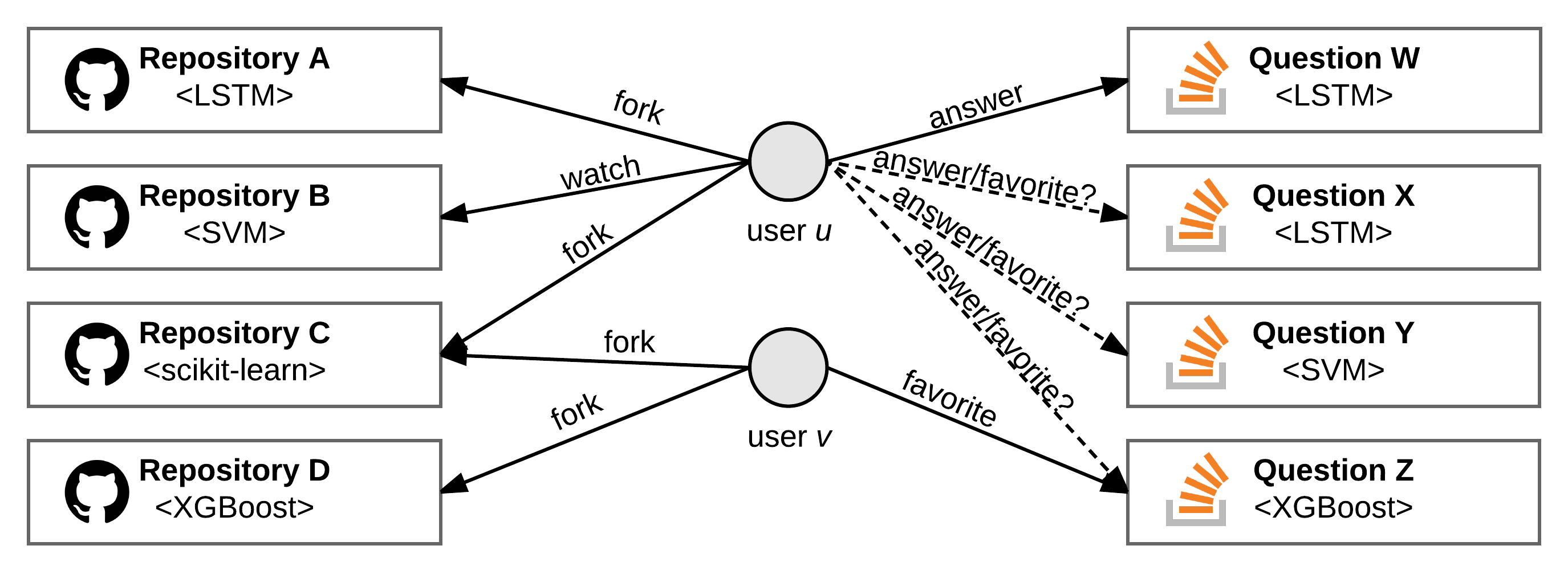}
	\caption{Example of Activity Prediction in Multi-Platform Setting}
	\label{fig:example}
\end{figure*}

Figure \ref{fig:example} illustrates an example for activity prediction in a multi-platform setting. Consider user $u$, who has accounts on both GitHub and Stack Overflow. If we adopt a \textit{direct platform activity prediction} approach, i.e., predicts a user's activities in a platform using his or her activity interests from the same platform, we could predict that \textit{u} is likely to answer or {\em favorite}\footnote{Bookmark a question in Stack Overflow} question \textit{X} in Stack Overflow as \textit{u} has previously answered a \textit{LSTM} related question. However, if we adopt a \textit{cross-platform activity prediction} approach, i.e., predicts a user's activities in a platform using his or her activity interests from another platform, we could predict that \textit{u} is also likely to answer or favorite a \textit{SVM} related question \textit{Y} as \textit{u} has previously \textit{watched}\footnote{Subscribe and receive updates on a repository in GitHub} a \textit{SVM} related repository \textit{B} in GitHub.  

The social collaborative nature of these platforms could also be exploited for activity prediction. According to Lee and Lo \cite{lee2017}, users who participated in the same GitHub repositories and Stack Overflow questions tend to share common interests. Therefore, we also explore the possibility to expand a user's interests to include the interests of users whom he or she had co-participated activities with. Referencing to the same example in Figure \ref{fig:example}, user \textit{v} co-forked the same \textit{scikit-learn} related repository as \textit{u} and forked another \textit{XGBoost} related repository. Although \textit{u} did not participate in any \textit{XGBoost} related repositories and questions, we can ``expand" \textit{u}'s interests to include \textit{XGBoost} as it is an interest of user \textit{v}. Finally, we could predict that \textit{u} is likely to also answer or favorite a \textit{XGBoost} related question. This expansion of interests could be particularly useful when a user has participated in very few activities on either platform. Note that the example also work for predicting GitHub activities using user's activities in Stack Overflow.

There are a number of benefits for using user interests from multi-platforms for activity prediction. Firstly, it enables prediction and recommendation of user activities in social collaborative platforms even when past activity history of a user is minimal or unavailable, i.e, cold-start problem \cite{schein2002}. For example, if we learn from a user's activities in GitHub that she is interested in \textit{Python} and \textit{text mining} techniques, we would predict that she will likely participate in \textit{Python} and \textit{text mining} related Stack Overflow questions even when she has just newly joined Stack Overflow and has not participated in any questions. Second, it could cover the {\em blind spots} of activity recommender systems which use only data from a single platform. For example, if a user has forked \textit{Android} related repositories in GitHub, recommender systems which are built on user's past activity in GitHub will likely to recommend the user more \textit{Android} related repositories. However, the same user may have also participated in some \textit{iOS} related questions in Stack Overflow, and such observations can be used to make relevant GitHub activity recommendations to the user.

\textbf{Contributions.} This work improves the state-of-the-art of inter-platform studies on multiple social collaborative platforms. Key contributions of this work include: Firstly, we proposed a novel framework which enables predicting users' activities using interests inferred from their activities in multiple social collaborative platforms. Secondly, we evaluate our method using large real-world datasets from Stack Overflow and GitHub. The results from our prediction experiments show that our proposed method is able to predict users' activities in GitHub and Stack Overflow with good accuracy, achieving an AUC score of up to 0.75 and 0.89 respectively.

\textbf{Paper outline.} The rest of the paper is organized as follows. Section \ref{sec:model} introduces the social collaborative platform activity prediction problem and describe our proposed the multi-platform prediction framework. Section \ref{sec:data} describe the data extraction process and the two real-world datasets, Stack Overflow and GitHub, that we used in our prediction experiments. Section \ref{sec:experiment} presents our experiments to predict user activities in the two social collaborative platforms using our proposed framework. Threat to validity of our study are discussed in Section \ref{sec:threats}. Section \ref{sec:related} reviews the literature related to our study. Finally, we summarize and conclude our work in Section \ref{sec:conclusion}.

%========================================================================================
\section{Proposed Method}
\label{sec:model}

In this section, we first present our proposed multi-platform activity prediction framework. We then define the prediction problem and describe the features used in our proposed prediction method.

\subsection{Multi-Platform Prediction Framework}
\begin{figure*}[h]
	\centering
	\includegraphics[scale = 0.65]{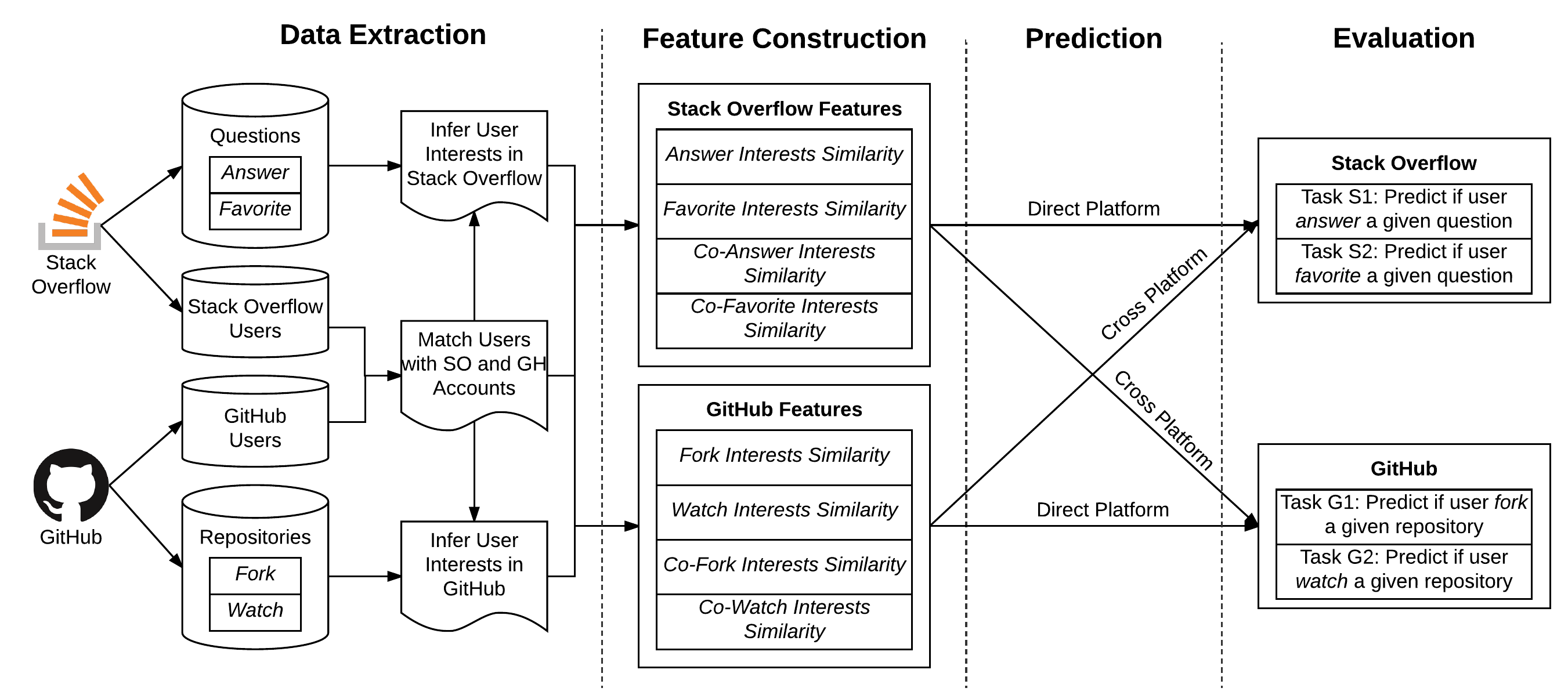}
	\caption{Cross-Platform Activity Prediction Framework}
	\label{fig:framework}
\end{figure*}

Figure \ref{fig:framework} shows the framework that we adopt for multi-platform activity prediction. We begin with data extraction from two social collaborative platforms: Stack Overflow and GitHub. There are three sub-processes in data extraction: (i) matching of users Stack Overflow and GitHub accounts, (ii) extracting the users' platform activities, and (iii) inferring users' interests from their activities. The details of these sub-processes will be covered in Section \ref{sec:data}. Next, we construct the Stack Overflow and GitHub user features which we will use in our prediction. 

Our framework also incorporates two approaches to predict users' platform activities, namely: \textit{direct} and \textit{cross} platform activity prediction. We define \textit{direct platform activity prediction} as predicting a user's platform activity using features from the same platform. For example, we predict if a given user will answer a given Stack Overflow question using the user's Stack Overflow features. Conversely, we define \textit{cross-platform activity prediction} as predicting a platform activity to a user using features from a different platform. For example, we predict if a given user will answer a given Stack Overflow question using the user's GitHub features. The performance of both prediction approaches will be evaluated on four prediction tasks, which will be described in Section \ref{sec:experiment}.

\subsection{Problem Statement}
Given a pair of query user and item (i.e., question or repository), $(u,k)$, we aim to predict if \textit{u} will perform an activity (e.g. answer, favorite, fork or watch) on \textit{k}. There are various ways to measure the likelihood of \textit{u} performing an activity on \textit{k}. For example, we could consider the similarity between \textit{k}'s description and \textit{u}'s interests inferred from different activities, or the similarity between \textit{k}'s description and the inferred interests of the user who co-participate activities with \textit{u}. In our proposed framework, we propose two types of user features, namely:\textit{user activity interest similarity features} and \textit{user co-activity interest similarity features}. The notations used throughout this paper are summarized in Table \ref{tab:notation}. 

We denote the estimated interests of a user given a repository $r$ that he or she forked and watched in GitHub as $I(r)$. Similarly, we denote the estimated interests of a user given a question $q$ that he or she answered and favorited in Stack Overflow as $I(q)$. Since the estimated interests given a repository or a question is the same for all users participated in it, we also refer to $I(r)$ and $I(q)$ as the interests in $r$ and $q$. For simplicity, we also refer to them as $r$'s interests and $q$'s interests respectively.

\begin{table}[h]
	\centering
	\caption{List of notations used}
	\label{tab:notation}
	\begin{tabular}{|l|p{6cm}|}
		\hline
		\textbf{Symbol} &  \textbf{Description}  \\ \hline
		$u$ & Query user\\
		$k$ & Query item\\
		$v$ & User who co-participated activities with user $u$\\
		$r$ & Repository\\
		$q$ & Question\\
		$I(r)$ & Interests of repository $r$\\
		$I(q)$ & Interests of question $q$\\
		$I(k)$ & Interests of query item $k$\\
		$u.RF$ & Set of repositories forked by user $u$\\
		$u.RW$ & Set of repositories watched by user $u$\\
		$u.QA$ & Set of questions answered by user $u$\\
		$u.QF$ & Set of questions favorited by user $u$\\
		$Co^{Fork}(u)$ & Set of users who co-forked at least one repository with user $u$\\
		$Co^{Watch}(u)$ & Set of users who co-watched at least one repository with user $u$\\
		$Co^{Ans}(u)$ & Set of users who co-answered at least one question with user $u$\\
		$Co^{Fav}(u)$ & Set of users who co-favorited at least one question with user $u$\\
		\hline
	\end{tabular}
\end{table}

\subsection{User Activity Interest Similarity Features}
This set of features measures the similarity between a query item $k$ and a query user $u$'s \textit{fork}, \textit{watch}, \textit{answer} and \textit{favorite} activity interests in GitHub and Stack Overflow. The intuition behind this set of features comes from the empirical study from Lee and Lo \cite{lee2017}, where they found that users in GitHub and Stack Overflow shared similarities between their interests in different types of activities and across the two platforms. Suppose that we want to predict if a user would fork a given repository in GitHub, we would measure the similarity between the given repository's interests and the developer's interests for the different activity types. Intuitively, the higher the similarity scores, the more likely the user would fork the given repositories. Equation \ref{simFork_eqn} captures the above intuition and measures similarity between $k$ and $u$'s fork activity interests (i.e., $Sim_{Fork}(u,k)$), by dividing $\{r\in u.RF|I(r)\in I(k))\}$, which is the number of $u$'s forked repositories that shared common interests with the item interests of $k$, by the total number of repositories forked by $u$ (i.e.,$u.RF$). 

\textbf{Example}. Referencing to the earlier example in Figure \ref{fig:example}, we could predict if user $u$ will answer question $X$ by computing the similarity between question $X$ and $u$'s fork activity interests. In this example, the common interests between $u$ and question $X$ will be \textit{LSTM}. The number of $u$'s forked repositories that shared common interests with question $X$ (i.e.,$\{r\in u.RF|I(r)\in I(k))\}$) will then be 1 (i.e., Repository $A$), while the total number of repositories forked by $u$ is 2 (i.e., Repository $A$ and $B$). Thus, $Sim_{Fork}(u,k) = \frac{1}{2} = 0.5$.

\begin{equation} \label{simFork_eqn}
Sim_{Fork}(u,k) = \frac{|\{r\in u.RF|I(r)\in I(k))\}|}{|u.RF|}
\end{equation}

\begin{equation} \label{simWatch_eqn}
Sim_{Watch}(u,k) = \frac{|\{r\in u.RW|I(r)\in I(k)\}|}{|u.RW|}
\end{equation}

\begin{equation} \label{simAns_eqn}
Sim_{Ans}(u,k) = \frac{|\{q\in u.QA|I(q)\in I(k)\}|}{|u.QA|}
\end{equation}

\begin{equation} \label{simFav_eqn}
Sim_{Fav}(u,k) = \frac{|\{q\in u.QF|I(q)\in I(k)\}|}{|u.QF|}
\end{equation}

We compute the similarities between $k$ and $u$'s watch, answer and favorite activities interests in similar ways as shown in Equation \ref{simWatch_eqn}, \ref{simAns_eqn} and \ref{simFav_eqn} respectively. 

\subsection{User Co-Activity Interest Similarity Features}
This set of features measures the similarity between a query item $k$ and the activity interests of other users $v$ who have co-participated in an activity with a query user $u$. The intuition behind this set of features also comes from the empirical study from Lee and Lo \cite{lee2017}, where they found that users share similar interests with other users who they co-participated an activity (even minimally) in a social collaborative platform. Suppose that we want to predict if a user would fork a given repository in GitHub, we would measure the similarity between the given repository's interests and the interests of other users who had co-forked repositories with the user in GitHub. Intuitively, we would also expect that the higher the similarity score, the more likely the user would answer the given question. Equation \ref{simCoFork_eqn} captures the above intuition and measures the average similarity between $k$ and fork activity interests of all users $v$, who had co-forked at least one question with $u$ (i.e., $Co^{Fork}(u)$).

As users also share common interests across different activities and platforms, we would expect that considering other users who had co-participated in other types of platform activities with the target user can also potentially help to predict if the target user would participate in a given platform activity. For instance, we are potentially able to predict if a user would fork a given repository by measuring the similarity between the given repository's interest and the interests of other users who have co-participated with the user in \textit{watch}, \textit{answer} and \textit{favorite} activities.

\textbf{Example}. Referencing to the example in Figure \ref{fig:example}, we could predict if user $u$ will favorite question $Z$ by computing the similarity between question $Z$ and the fork activity interests of other users who have co-fork a repository with user $u$. Assuming that user $u$ only has 1 other user, $v$, who co-fork repositories with him or her, the common interests between $v$ and question $Z$ will be \textit{XGBoost}. The number of $v$'s forked repositories that shared common interests with question $Z$ (i.e.,$\{r\in v.RF|I(r)\in I(k))\}$) will then be 1 (i.e., Repository $C$), while the total number of repositories forked by $v$ is 2 (i.e., Repository $C$ and $D$). Finally, $Sim_{CoFork}(u,k) = \frac{\frac{1}{2}}{1} = 0.5$. 

\begin{equation} \label{simCoFork_eqn}
Sim_{CoFork}(u,k) = \frac{\bigg[\sum_{v\in Co^{Fork}(u)} \frac{|\{r\in v.RF|I(r)\in I(k)\}|}{|v.RF|}\bigg]}{|Co^{Fork}(u)|} 
\end{equation}

\begin{equation} \label{simCoWatch_eqn}
Sim_{CoWatch}(u,k) = \frac{\bigg[\sum_{v\in Co^{Watch}(u)} \frac{|\{r\in v.RW|I(r)\in I(k)\}|}{|v.RW|}\bigg]}{|Co^{Watch}(u)|} 
\end{equation}

\begin{equation} \label{simCoAns_eqn}
Sim_{CoAns}(u,k) = \frac{\bigg[\sum_{v\in Co^{Ans}(u)} \frac{|\{q\in v.QA|I(q)\in I(k)\}|}{|v.QA|}\bigg]}{|Co^{Ans}(u)|} 
\end{equation}

\begin{equation} \label{simCoFav_eqn}
Sim_{CoFav}(u,k) = \frac{\bigg[\sum_{v\in Co^{Fav}(u)} \frac{|\{q\in v.QF|I(q)\in I(k)\}|}{|v.QF|}\bigg]}{|Co^{Fav}(u)|}  
\end{equation}

We compute the similarities between $k$ and activity interests of other users $v$ who have co-watched, co-answered and co-favorited with a target user $u$ in similar ways as shown in Equation \ref{simCoWatch_eqn}, \ref{simCoAns_eqn} and \ref{simCoFav_eqn} respectively.

\section{Data Extraction \& Examination}
\label{sec:data}
In this section, we first introduce the two large real-world datasets that we use in our activity prediction experiments. Next, we discuss the user accounts linkage process to retrieve users who are active in multiple social collaborative sites, and a summary of the users' activities retrieved. We then discuss the heuristic used to infer user interests from their participated activities. Finally, we empirically examine the similarity between the GitHub repositories and Stack Overflow questions participated by users on both social collaborative sites.  

\subsection{Datasets}
There are two main datasets used in our study. For the GitHub dataset, we use the MongoDB database dump released on March 2015~\cite{Gousi13}. The dataset contains GitHub activities from October 2013 to March 2015 of about 2.5 million users. Specifically, we are interested in the \textit{fork} and \textit{watch} repositories activities of the GitHub users. For Stack Overflow, we use the XML dataset released on March 2015\footnote{https://archive.org/details/stackexchange}. This dataset contains information of estimated 1 million Stack Overflow users and their activities from October 2013 to March 2015. We are particularly interested in the \textit{answer} and \textit{favorite} activities of the Stack Overflow users.

\subsection{User Account Linkage}
As this study intends to investigate user interests across GitHub and Stack Overflow, we need to identify users who were using both platforms. For this work, we used the dataset provided by Badashian et al. \cite{badashian2014}, where they utilized GitHub users' email addresses and Stack Overflow users' email MD5 hashes to find the intersection between the two datasets. We also filter out users who do not have at least 1 activity on both platforms between October 2013 and March 2015. In total, we identify 92,427 users, which forms our \textit{base users} set. After the base users have been identified, we extract their GitHub and Stack Overflow activities from the datasets. In total, we have extracted 416,171 \textit{fork}, 2,168,871 \textit{watch}, 766,315 \textit{answer} and 427,093 \textit{favorite} activities from the base users.

\begin{figure}[h]
	\centering
	\includegraphics[scale = 0.14]{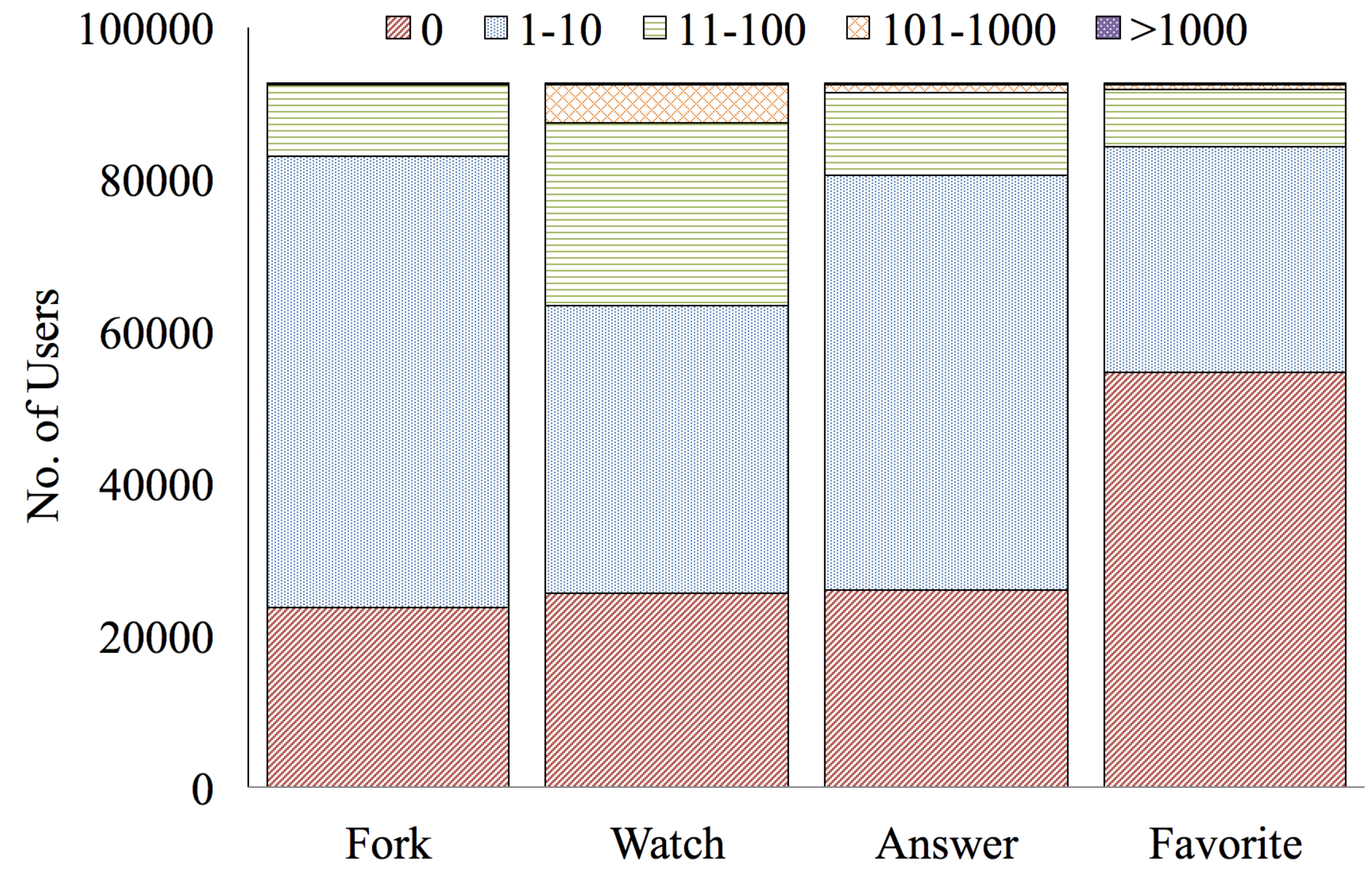}
	\caption{Base users' Stack Overflow and GitHub Activity Distributions}
	\label{fig:dataset}
\end{figure}

Figure \ref{fig:dataset} shows the distributions of base users' activities in GitHub and Stack Overflow. Most of the base users forked 1-10 repositories (64\% of the base users), answered 1-10 questions (54\% of the base users). There are also quite a number of developers who watched 11-100 repositories (26\% of the base users). We also observe that more than half of base users have at least answered 1 questions (71\%) and a substantial number of base users also answered 11-100 questions (12\%).Interestingly, the high contribution of answers to questions in Stack Overflow could also suggest that the many of these active developers in our study were experts in their domain or areas of interest. Lastly, we also notice that there are developers (albeit very few in number) who were extremely active in GitHub and Stack Overflow; they forked, watched, committed, pull-requested more than 1000 repositories, or asked, answered and favorited more than 1000 questions.

\subsection{Inferring User Interests}
Next, we infer user interests by observing repositories and questions that users participated in GitHub and Stack Overflow. We use the following heuristics to infer user interests:

\begin{enumerate}
    \item To infer user interests in Stack Overflow, we use the descriptive tags of the questions that they answered and favorited. For example, consider a question related to mobile programming for Android smartphones which contain the following set of descriptive tags: \{\textit{Java, Android}\}. If a user answered, or favorited that question, we infer that his interests include \textit{Java} and \textit{Android}.
    
    \item In the time period covered in our dataset, GitHub does not allow users to tag repositories but it allows users to describe their repositories. These descriptions often contain important keywords that can shed light on user interests. To infer user interests from the repositories that a user had participated, we first collect all descriptive tags that appear in our Stack Overflow dataset. In total, 39,837 unique descriptive tags are collected. Next, we perform keyword matching between the collected Stack Overflow tags and a GitHub repository description. We consider the matched keywords as the inferred interests. We choose to use Stack Overflow tags to ensure that developer interests across the two platforms can be mapped to the same vocabulary.
\end{enumerate}

\subsection{Similarity between GitHub repositories and Stack Overflow Questions}
In \cite{lee2017}, Lee and Lo had empirically studied the similarity between user activity interests in GitHub and Stack Overflow. We extend their study by examining the descriptive tags of GitHub repositories and Stack Overflow questions participated by the base users. The objective is to investigate what are the popular descriptive tags used by the users on two sites and if there are overlaps among the popular descriptive tags.

\begin{table}[h]
\centering
\caption{Top 10 most used descriptive tags in Stack Overflow and GitHub}
\label{tab:tags}
\begin{tabular}{c|l|c||l|c|}
\cline{2-5}
\multicolumn{1}{l|}{}      & \multicolumn{2}{c||}{Stack Overflow}     & \multicolumn{2}{c|}{GitHub}                \\ \hline
\multicolumn{1}{|c|}{Rank} & \multicolumn{1}{c|}{Tag} & \% Questions & \multicolumn{1}{c|}{Tag} & \% Repositories \\ \hline
\multicolumn{1}{|c|}{1}    & javascript              & 3.758        & javascript              & 8.163           \\ \hline
\multicolumn{1}{|c|}{2}    & java                    & 3.104        & ruby                    & 2.698           \\ \hline
\multicolumn{1}{|c|}{3}    & python                  & 2.760        & python                  & 2.604           \\ \hline
\multicolumn{1}{|c|}{4}    & c                       & 1.940        & c                       & 2.163           \\ \hline
\multicolumn{1}{|c|}{5}    & php                     & 1.884        & java                    & 1.928           \\ \hline
\multicolumn{1}{|c|}{6}    & android                  & 1.787        & objective-c              & 1.731           \\ \hline
\multicolumn{1}{|c|}{7}    & jquery                   & 1.577        & php                     & 1.330           \\ \hline
\multicolumn{1}{|c|}{8}    & ios                      & 1.459        & go                       & 1.275           \\ \hline
\multicolumn{1}{|c|}{9}    & ruby                    & 1.031        & css                     & 1.236           \\ \hline
\multicolumn{1}{|c|}{10}   & css                     & 0.963        & shell                    & 0.884           \\ \hline
\end{tabular}
\end{table}

Table \ref{tab:tags} shows the top 10 most used descriptive tags and the percentages of Stack Overflow questions and GitHub repositories containing these tags. We observe quite a significant number of overlap in the top 10 descriptive tags between the two social collaborative sites (known as \textit{overlapped tags}); i.e., \textit{javascript}, \textit{java}, \textit{python}, \textit{c}, \textit{php}, \textit{ruby} and \textit{css}. This suggests that generally, the base users participated in questions and repositories of similar domain and nature in Stack Overflow and GitHub. Another interesting observation is the proportions of questions and repositories with the top 10 descriptive tags; the overlapped tags are used in similar proportions of Stack Overflow questions and GitHub repositories, with the exception of \textit{javascript}, which seems to be more popular in GitHub (i.e., $\sim$8\% of repositories) than Stack Overflow (i.e., $\sim$3\% of questions).  

\begin{figure*}[t]
	\centering
	\setlength{\tabcolsep}{0pt} % Default value: 6pt
	\renewcommand{\arraystretch}{0} % Default value: 1
	\begin{tabular}{ccc}
		\includegraphics[scale = 0.5]{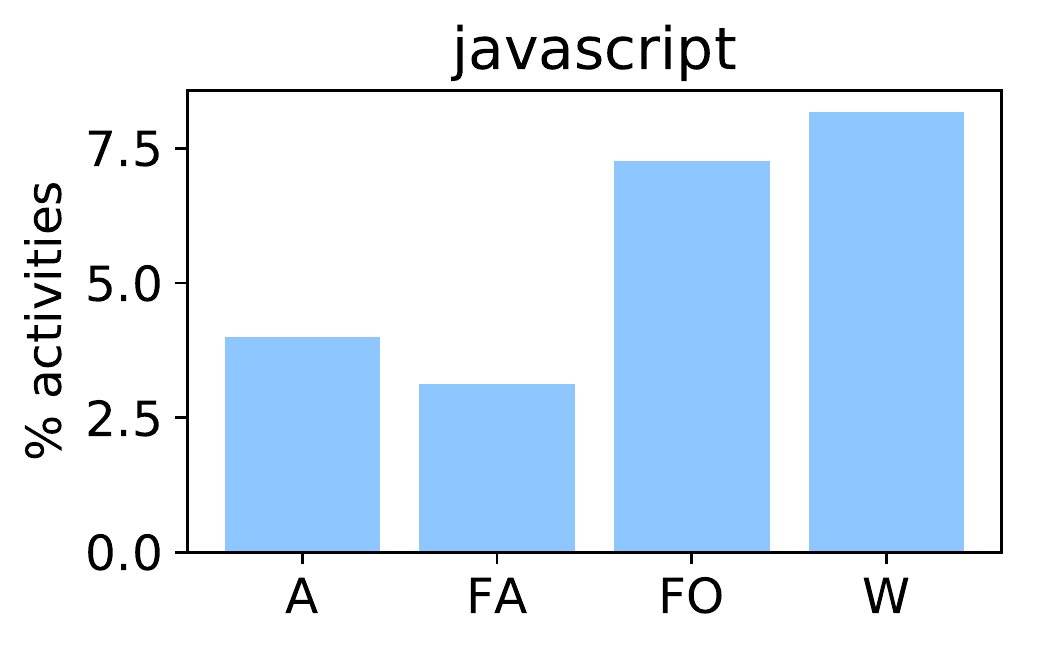} &   
        \includegraphics[scale = 0.5]{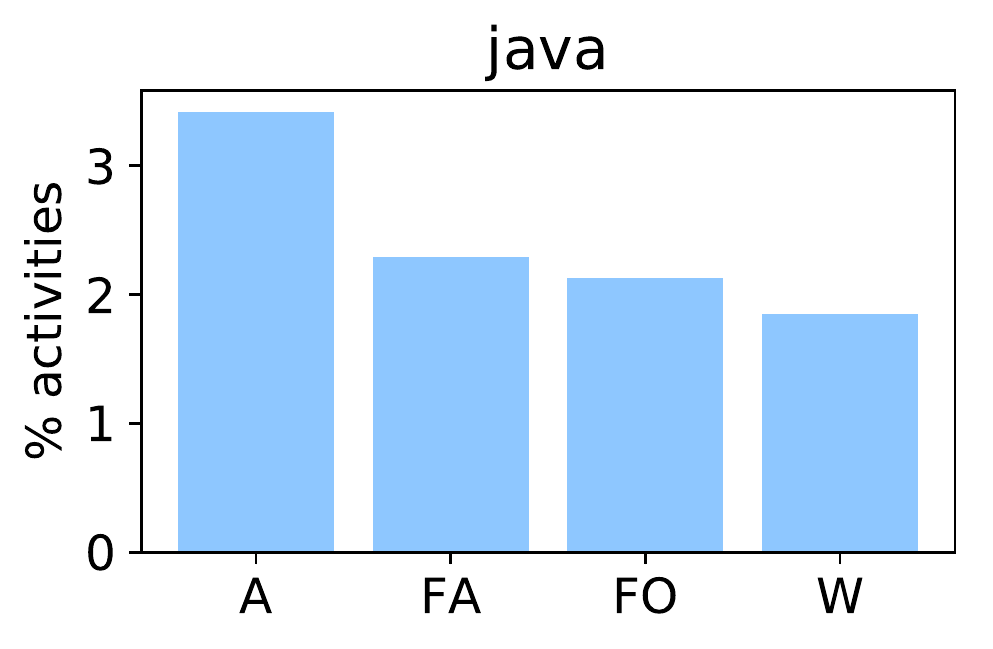} &
        \includegraphics[scale = 0.5]{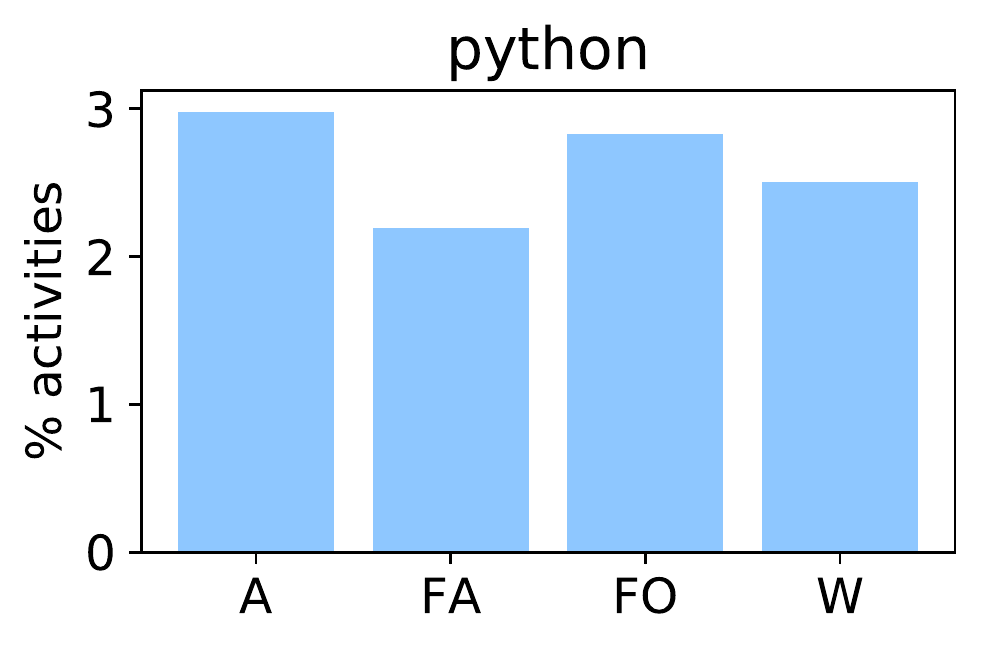} \\
        \includegraphics[scale = 0.5]{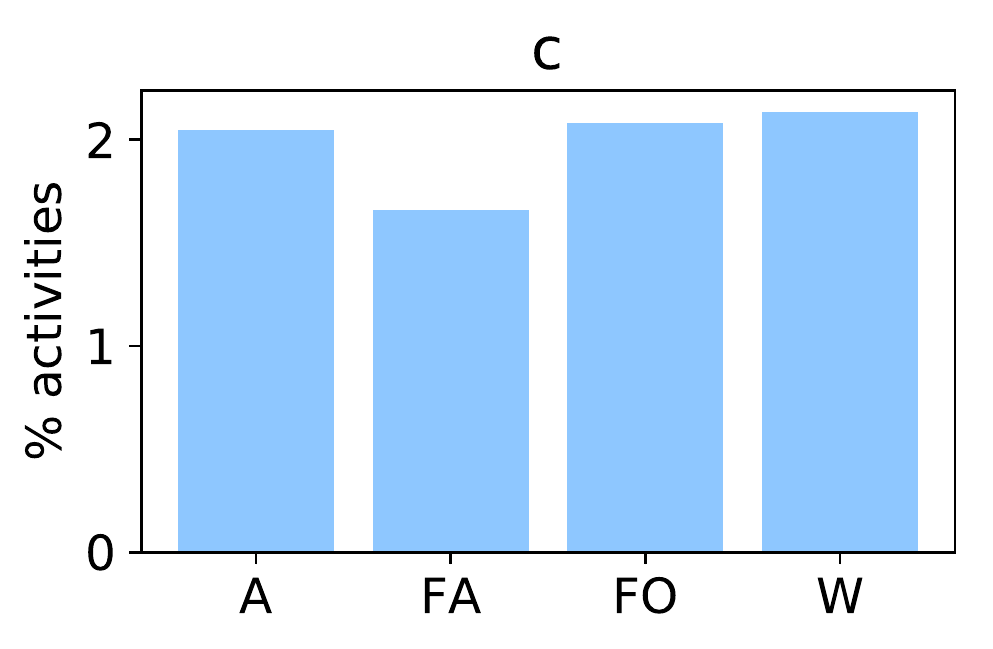} &
		\includegraphics[scale = 0.5]{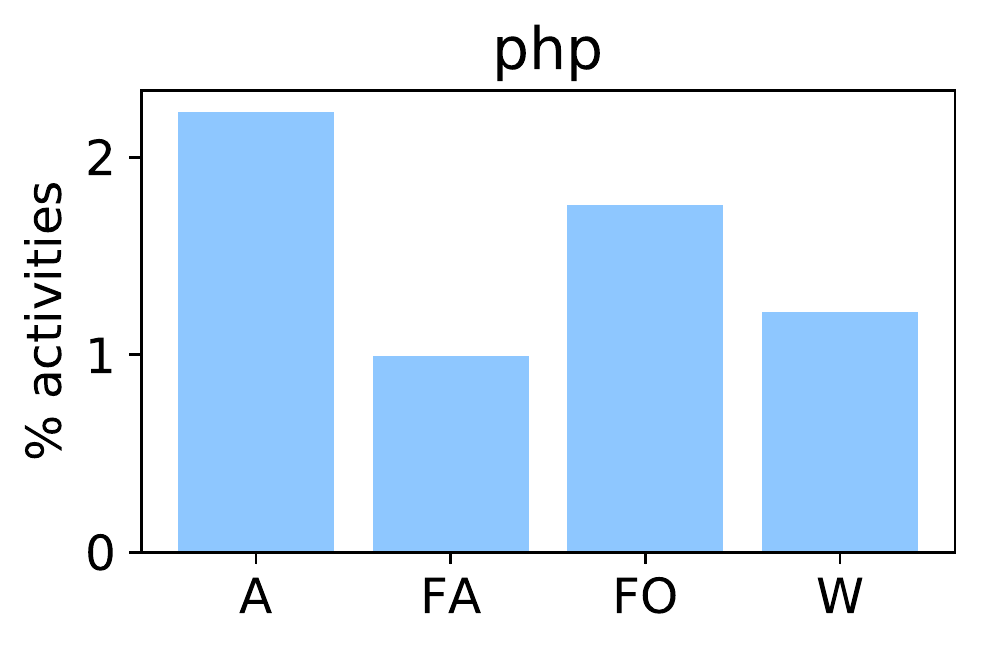} &   
        \includegraphics[scale = 0.5]{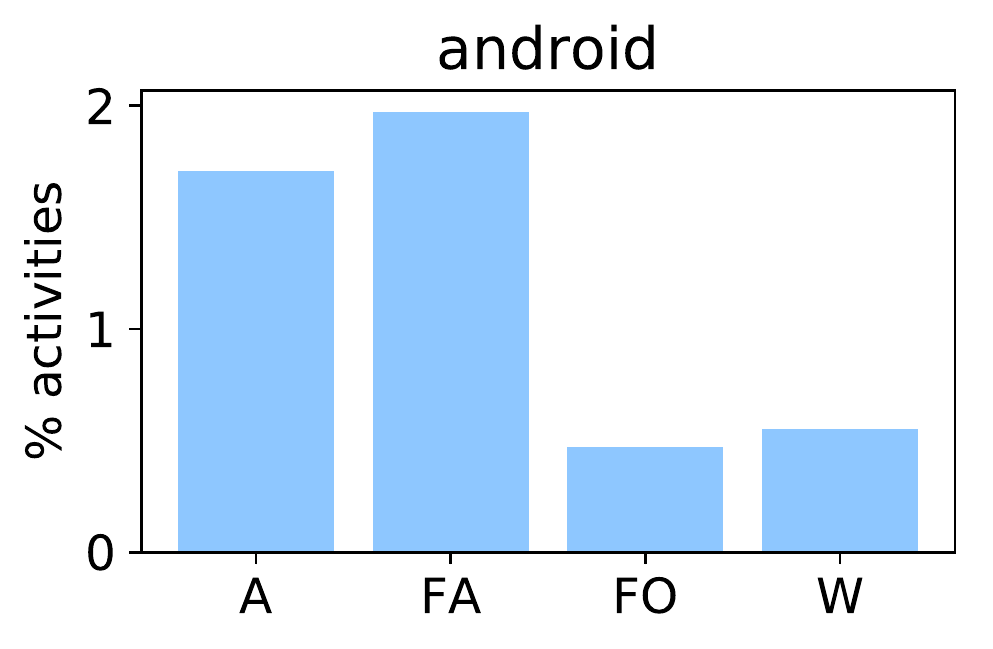} \\
        \includegraphics[scale = 0.5]{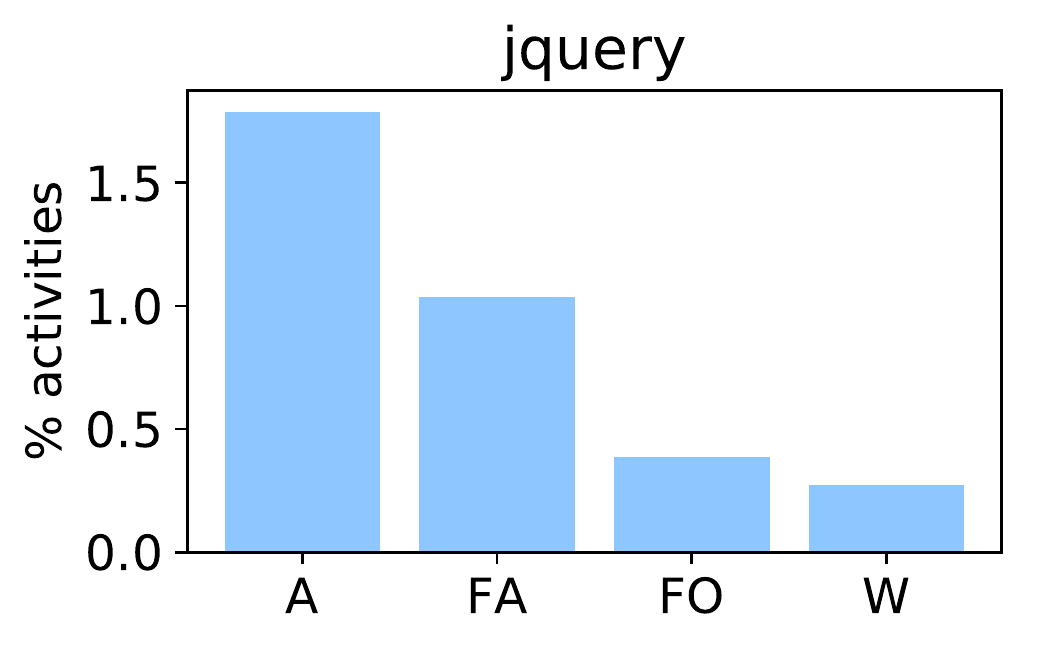} &
        \includegraphics[scale = 0.5]{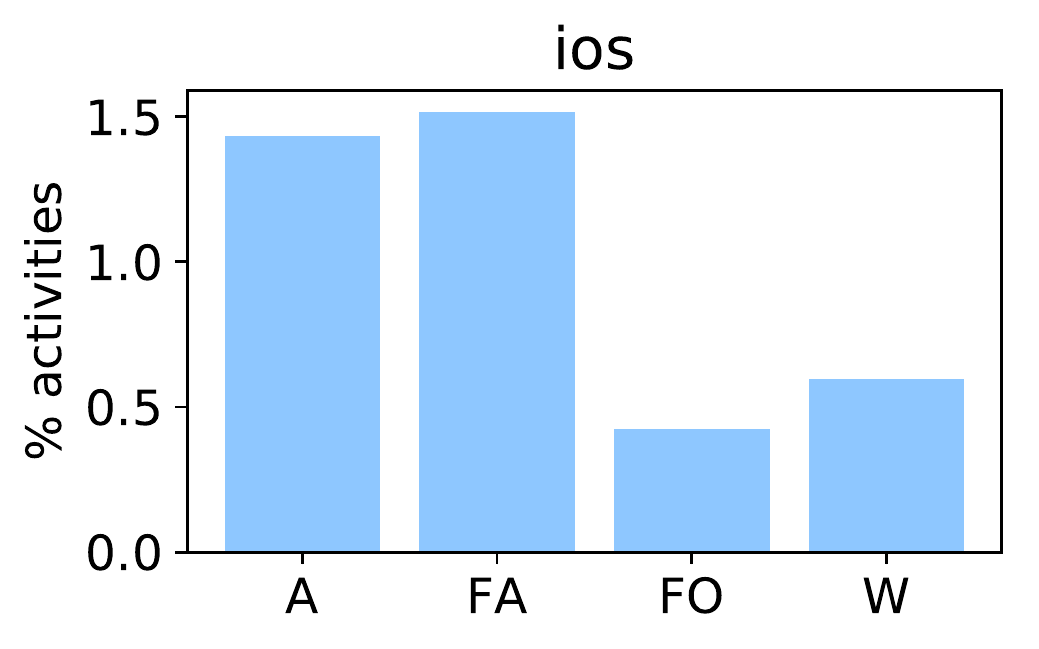} &
        \includegraphics[scale = 0.5]{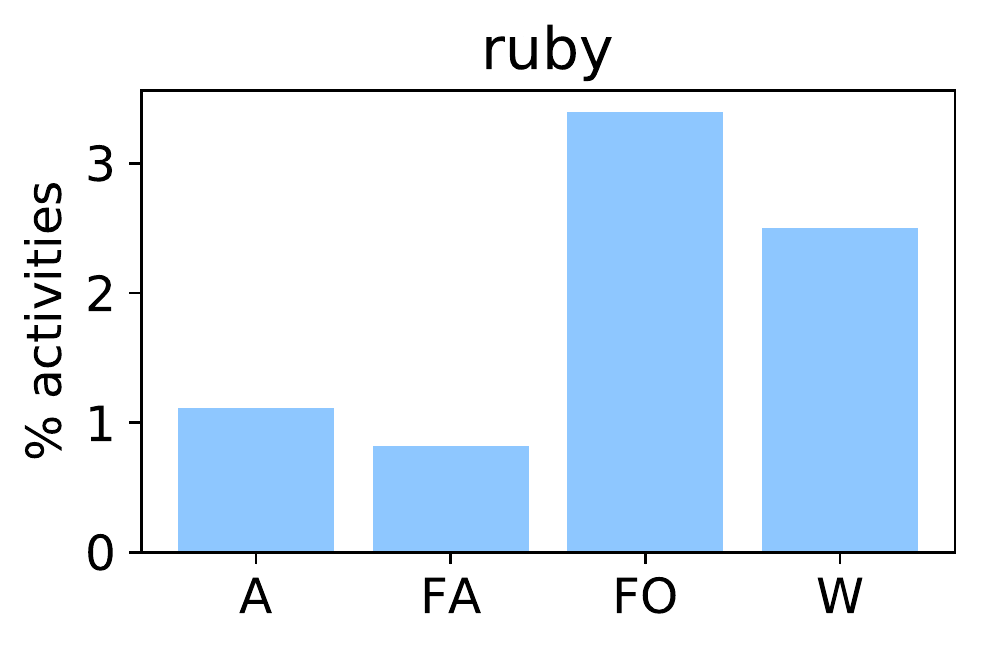} \\   
        \includegraphics[scale = 0.5]{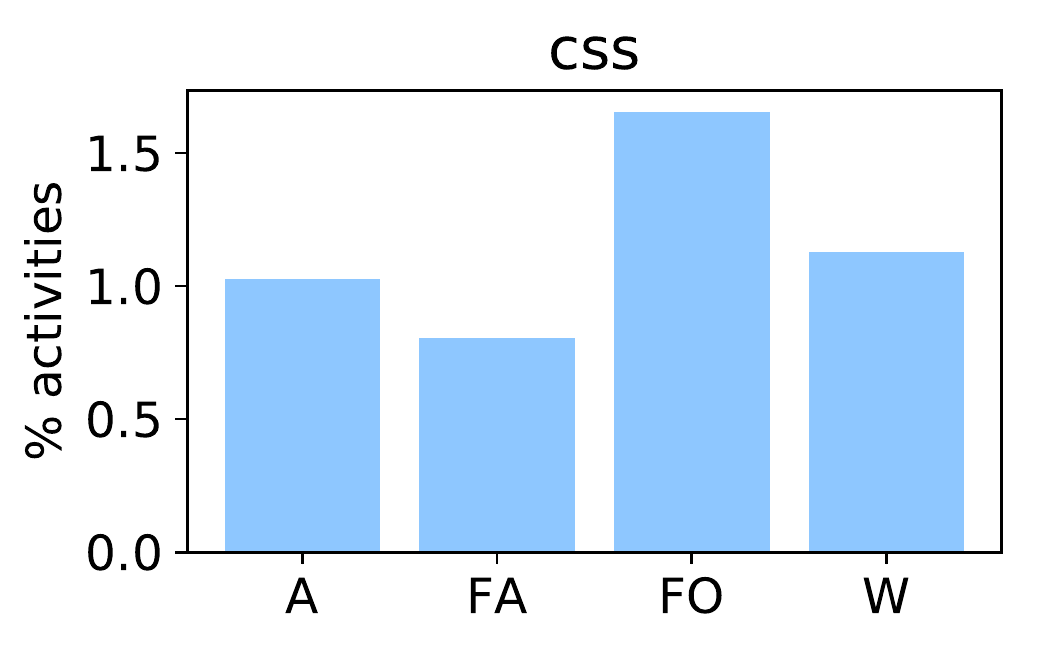} &
        \includegraphics[scale = 0.5]{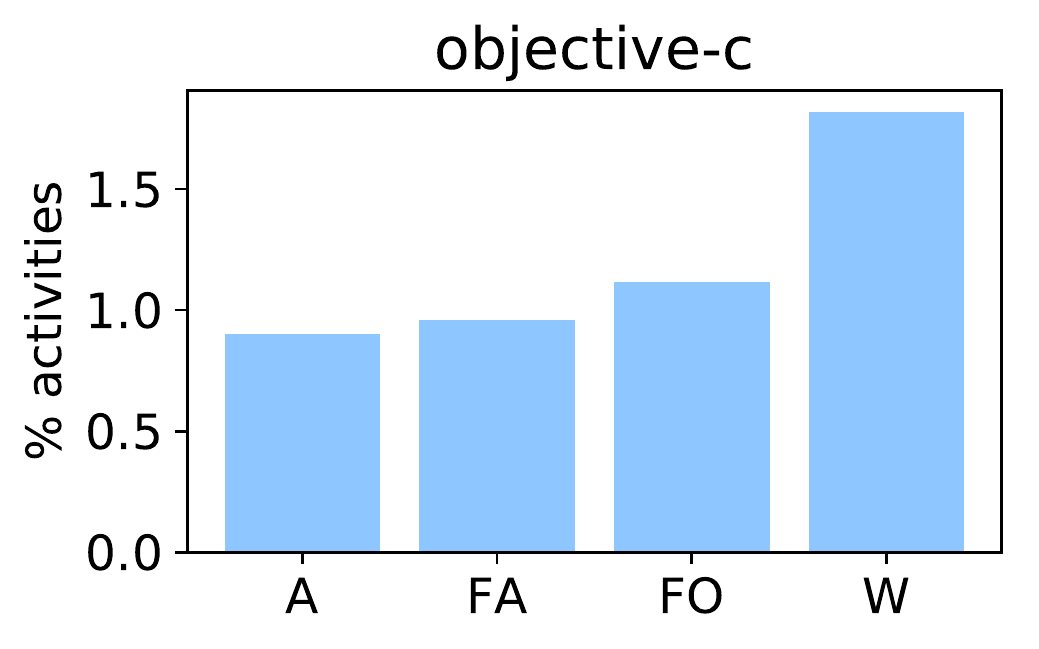} &
        \includegraphics[scale = 0.5]{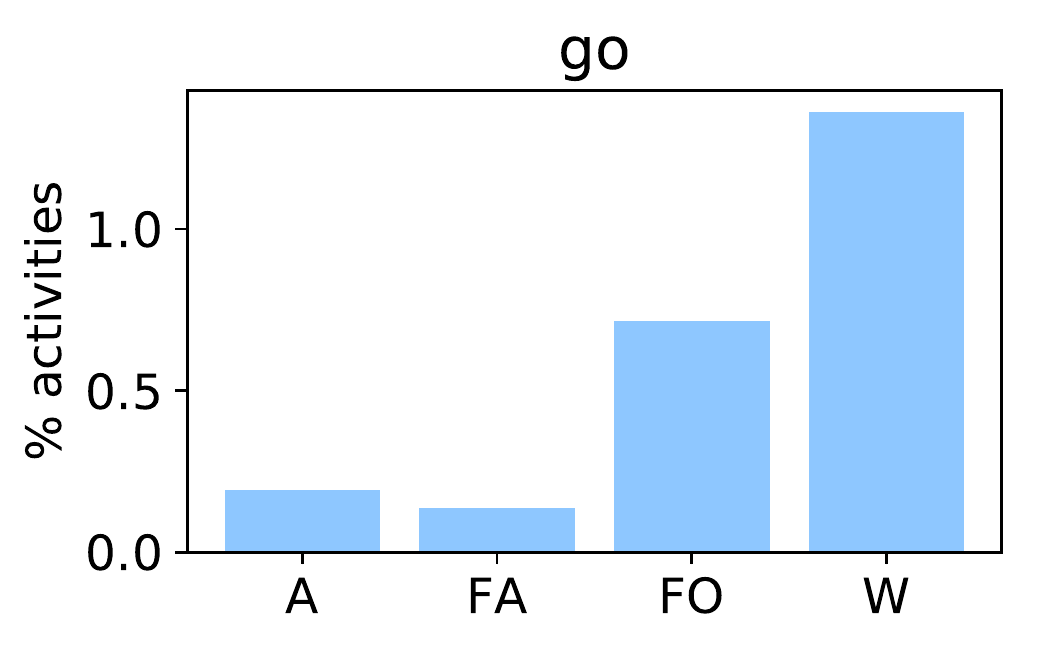} \\
        \multicolumn{3}{c}{\includegraphics[scale = 0.5]{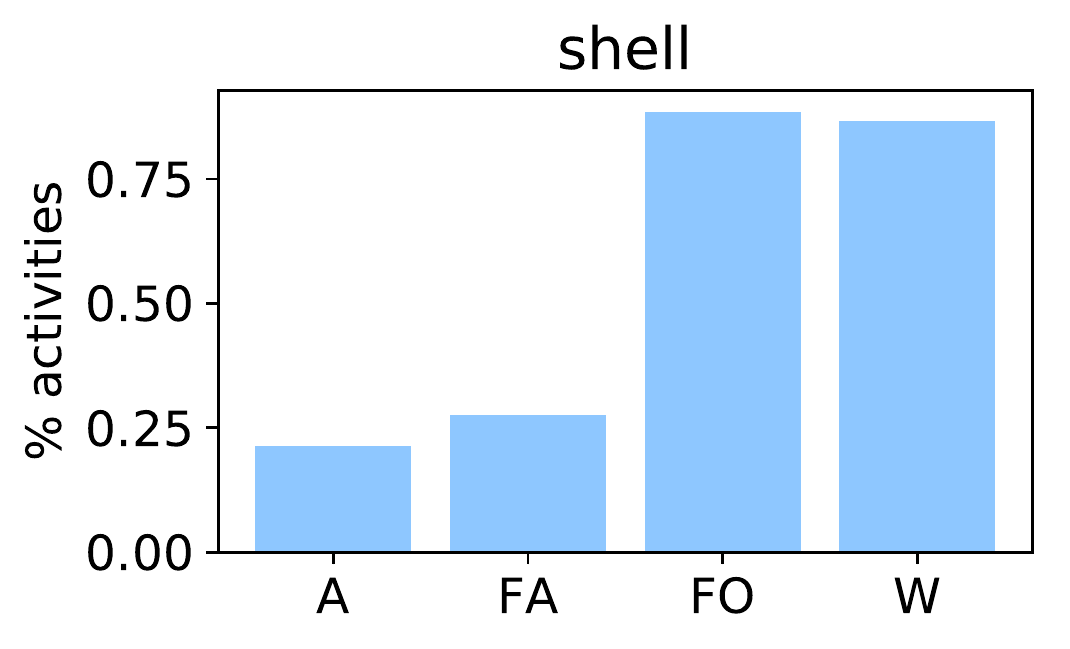}}
	\end{tabular}
	\caption{Bar chats of percentages of activities that use the different descriptive tags. Note that "A", "FA", "FO" and "W" denote \textit{answer}, \textit{favorite}, \textit{fork} and \textit{watch} activities respectively.}
	\label{fig:activitytags}
\end{figure*}

We further investigate the usage of these descriptive tags in the different Stack Overflow and GitHub user activities. Figure \ref{fig:activitytags} shows the activity bar charts for the most used descriptive tags. For example, $\sim$7\% of the repositories watched by the base users contained the descriptive tag \textit{javascript}, while only $\sim$3\% of the questions answered by the base users contained the same tag. We observe that for a given descriptive tag, the proportion of activities involving the tag is not uniform even within the same social collaborative site. For example, the base users answered more Stack Overflow questions on \textit{php} than favoriting them. Conversely, for \textit{android} related questions, the base users favorited these question more than answering them. This unevenness in activity proportions is also observed to be greater for activities involving non-overlaps descriptive tags. For example, the base users fork and watch significantly more \textit{shell} related repositories than answering and favoriting t\textit{shell} related questions. This suggests that although the users do participate in questions and repositories of similar domains, the activity preferences involving the similar domains varies. Thus, it would be more natural to learn the user interests at the activity level instead of aggregating the interests at the platform level.

%========================================================================================
\section{Experiments}
\label{sec:experiment}
In this section, we describe the supervised prediction experiments conducted to evaluate our proposed method. Specifically, we consider the following activity prediction tasks:

\begin{itemize}
	\item \textit{Answer Prediction}. Given a Stack Overflow \textit{user-question} pair, predict if the user will answer the question
	\item \textit{Favorite Prediction}. Given a Stack Overflow \textit{user-question}, predict if the user will favorite the question
	\item \textit{Fork Prediction}. Given a GitHub \textit{user-repository}, predict if the user will fork the repository
	\item \textit{Watch Prediction}. Given a GitHub \textit{user-repository}, predict if the user will watch the repository
\end{itemize}

\subsection{Experiment Setup}
\textbf{Data Selection.} For \textit{answer prediction} task, we retrieve all the Stack Overflow questions that the base users have answered and define a positive instance as a \textit{user-question} pair where a base user had answered the particular question in Stack Overflow. For negative instances, we randomly assign a Stack Overflow question to the base users and check that the randomly assigned pair does not exist in the positive instance set. For the training datasets used in \textit{answer prediction task}, we randomly generated 5,000 negative instances and randomly selected 5,000 positive instances from the questions answered by users between October 2013 and June 2014 (9 months). The same approach was used to generate the positive and negative instances for test sets using the questions answered by the users between July 2014 and March 2015 (9 months). Similar approach was used to generate the \textit{user-question} and \textit{user-repository} pairs for positive and negative instances used in \textit{favorite, fork} and \textit{watch prediction} tasks. 

Note that we have repeated the prediction experiments for five runs, and the random selection of train and tests set are repeated for each of the runs. Also, although we know the true labels of the \textit{user-question} and \textit{user-repository} pairs, we do not take the labels into consideration when deriving the values of our proposed features, i.e., we assume that we do not know the labels of the pairs. 

\textbf{Feature Configuration.} To compare the performance of \textit{direct} and \textit{cross} platform activity prediction approaches, we use Support Vector Machine (SVM) with linear kernel and apply the following feature sets on all prediction tasks:

\begin{figure*}[t]
	\setlength{\tabcolsep}{0pt} % Default value: 6pt
	\renewcommand{\arraystretch}{0} % Default value: 1
	\begin{tabular}{cccc}
		\includegraphics[scale = 0.38]{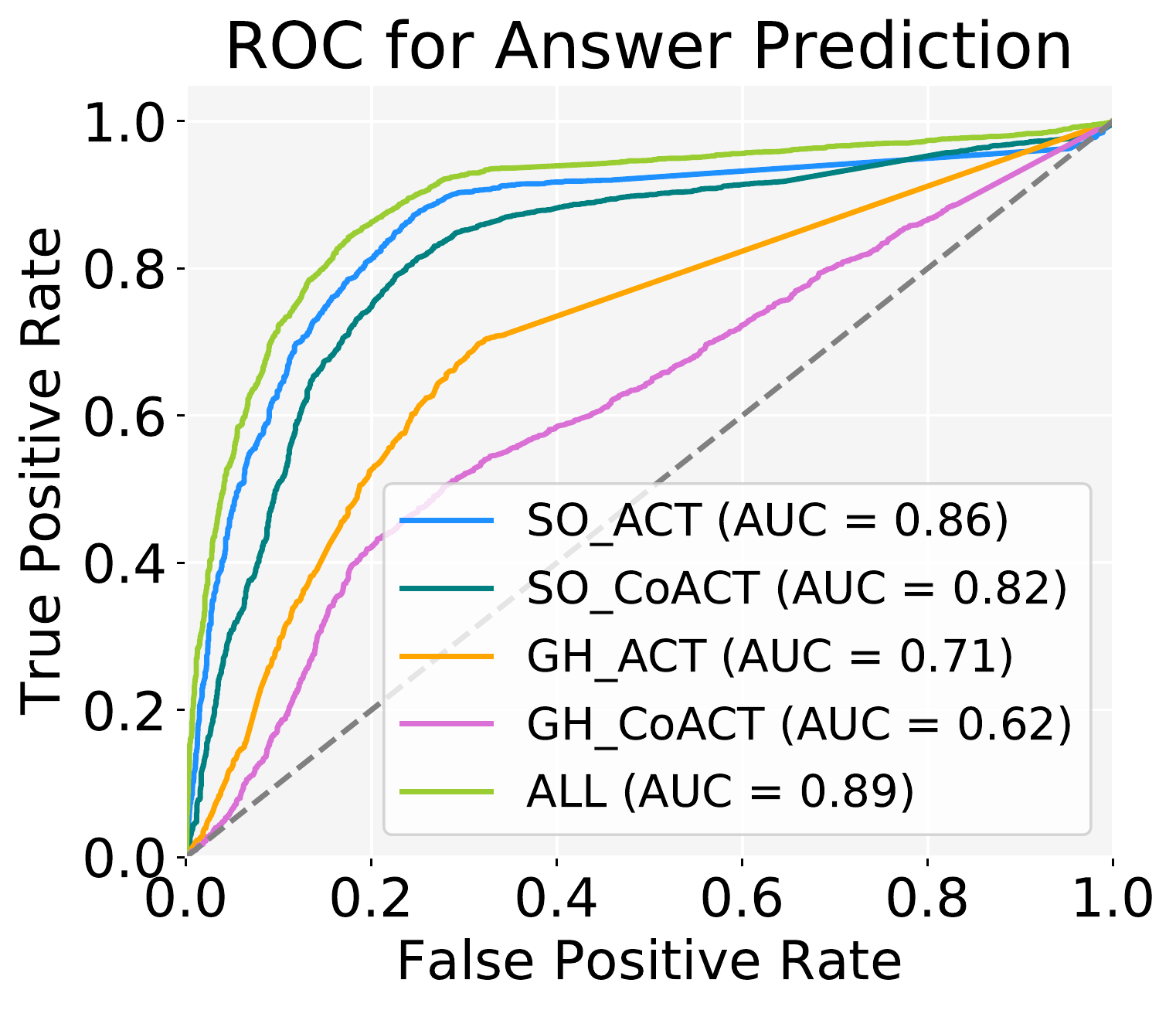} &   
        \includegraphics[scale = 0.38]{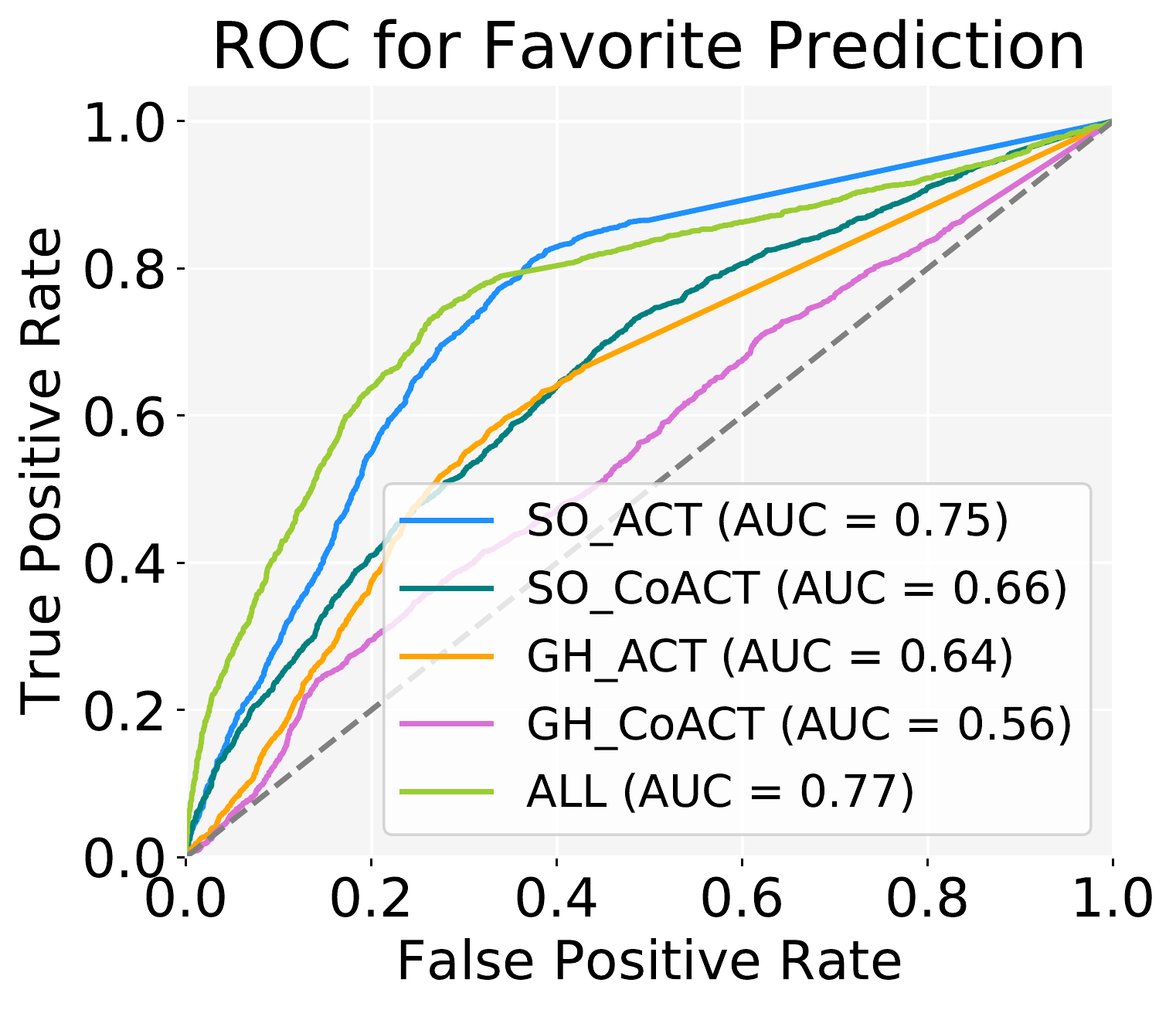}\\
		\includegraphics[scale = 0.38]{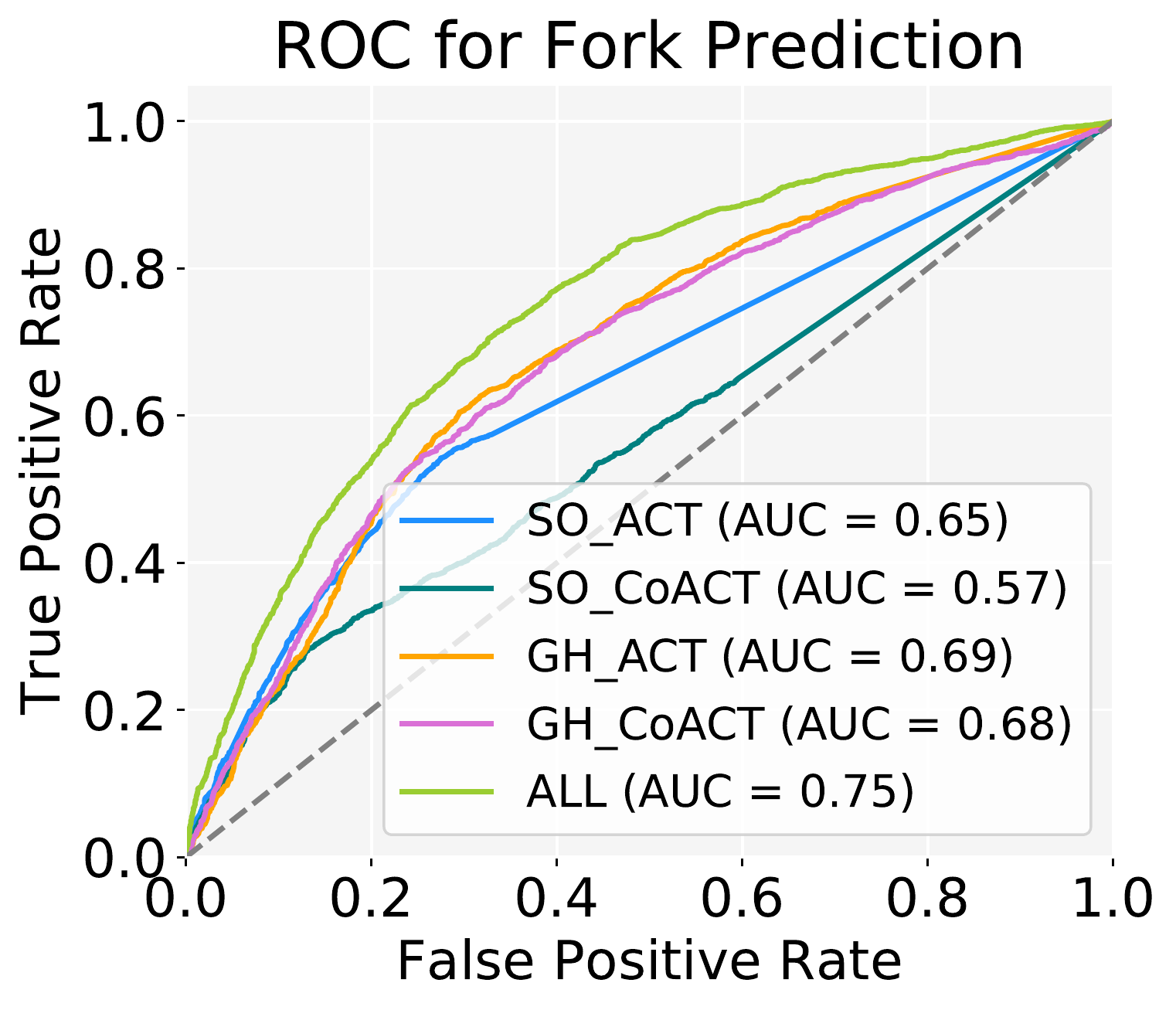} &   
        \includegraphics[scale = 0.38]{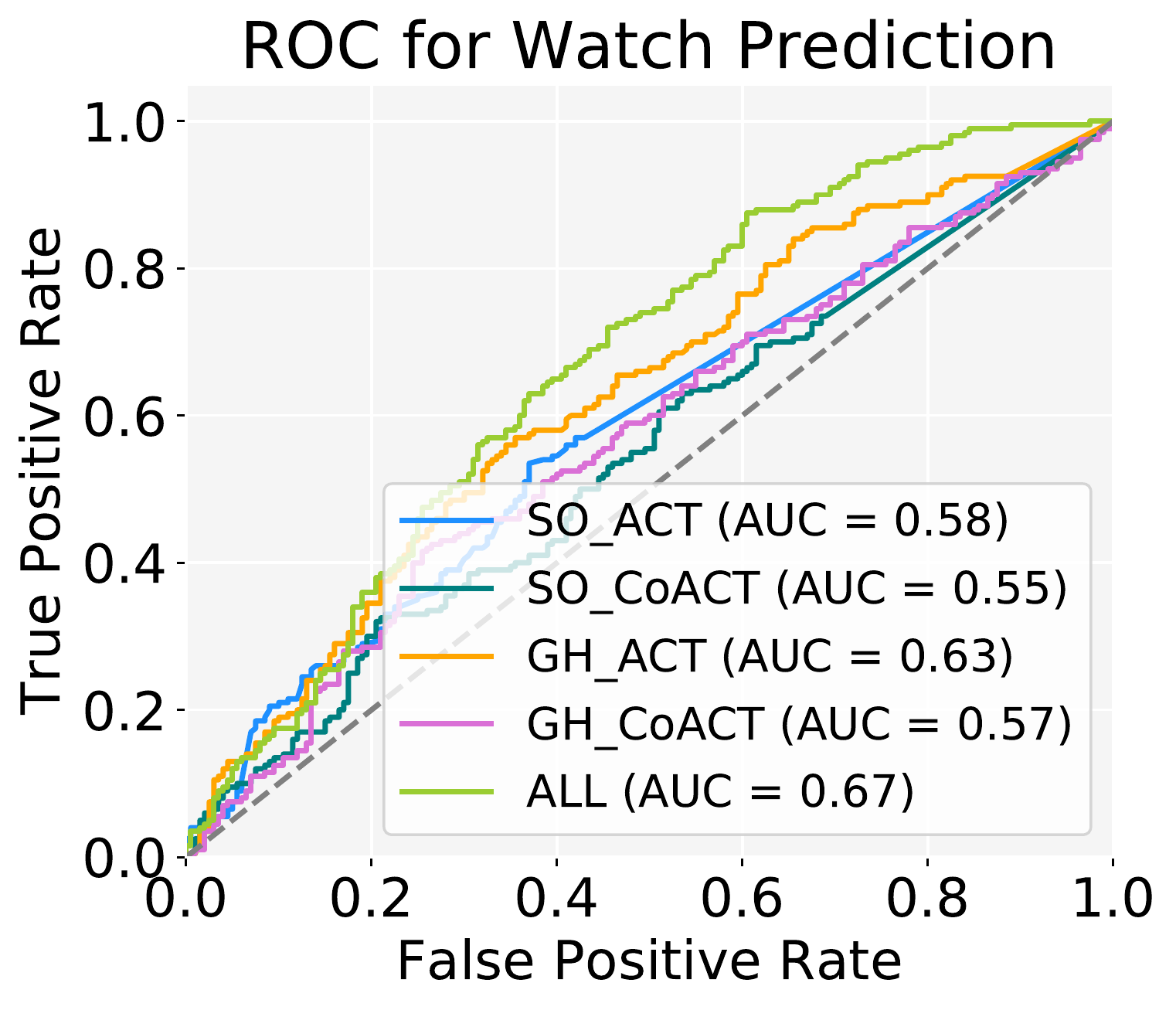} \\
	\end{tabular}
	\caption{ROCs for Four Prediction Tasks}
	\label{fig:results}
\end{figure*}

\begin{itemize}
	\item \textbf{SO\_Act}: This set of features includes the \textit{Answer} (Eqn. \ref{simAns_eqn}) and \textit{Favorite} (Eqn. \ref{simFav_eqn}) \textit{Interests Similarity} scores for a given user-question or user-repository pair. 
	\item \textbf{SO\_CoAct}: This set of features includes the \textit{Co-Answer} (Eqn. \ref{simCoAns_eqn}) and \textit{Co-Favorite} (Eqn. \ref{simCoFav_eqn}) \textit{Interests Similarity}  scores for a given user-question or user-repository pair.
	\item \textbf{GH\_Act}: This set of features includes the \textit{Fork} (Eqn. \ref{simFork_eqn}) and \textit{Watch} (Eqn. \ref{simWatch_eqn}) \textit{Interests Similarity} scores for a given user-question or user-repository pair. 
	\item \textbf{GH\_CoAct}: This set of features includes the \textit{Co-Fork} (Eqn. \ref{simCoFork_eqn}) and \textit{Co-Watch} (Eqn. \ref{simCoWatch_eqn}) \textit{Interests Similarity} scores for a given user-question or user-repository pair. 
	\item \textbf{ALL}: This set of features is the union of all features. 
\end{itemize}

\subsection{Prediction Results}
We measure the prediction accuracy for each feature configuration by computing the average area under the ROC curve (AUC) over a set of positive and negative examples drawn from the test set for each of the five runs. The results for the four prediction tasks are shown in Figure \ref{fig:results}. We observe that feature configuration {\bf ALL} performed the best in all prediction tasks, achieving an AUC of 0.89, 0.77, 0.75 and 0.67 for \textit{answer}, \textit{favorite}, \textit{fork} and \textit{watch} prediction tasks respectively. 

\textbf{Performance of cross-platform prediction approach.} Although the \textit{cross-platform prediction approach} did not outperform the \textit{direct platform prediction approach} in user activity prediction, they still yield good accuracy. For example, when predicting user's \textit{answer} and \textit{favorite} activities in Stack Overflow, the GitHub \textit{user activity interests similarity} features (i.e., \textbf{GH\_Act}) has AUC of 0.71 and 0.64 respectively, and when predicting user's \textit{fork} and \textit{watch} activities in GitHub, the Stack Overflow \textit{user activity interests similarity} features (i.e., \textbf{SO\_Act}) has AUC of 0.65 and 0.58 respectively.  The AUC for predicting user's answer activities in Stack Overflow using \textit{user activity interests similarity} features (i.e., \textbf{GH\_Act}) is observed to be slightly higher than the prediction for other activities. A possible explanation for this could be the difference between the nature of user activities; answering a question in Stack Overflow would require that a user possesses a particular domain expertise, whereas other activities such as watching a GitHub repository or favoriting a Stack Overflow question depend on the user's interests. As such, we observe higher AUC score for \textit{predicting answer activity} task as the users' expertise are usually more specialized and less diverse than their interests.

More interestingly, using \textit{cross-platform prediction approach} with \textit{user co-activity interests similarity} features (i.e., \textbf{GH\_CoAct} and \textbf{SO\_CoAct}), have also yielded reasonable prediction accuracies. For example, when predicting user's answer activities in Stack Overflow, \textbf{GH\_CoAct} has yielded an AUC of 0.62. This suggests that even with no information about a user's past activities in the Stack Overflow and only minimal information such as the user's co-activities in GitHub, we are still able to reasonably predict user's activity in Stack Overflow. Similar observations are made when predicting user activities in GitHub using user's co-activities in Stack Overflow. 

\subsection{Discussion}
The results of the four prediction tasks offer us some insights in performing recommendations in social collaborative platforms.

\textbf{Solving cold-starts.} The reasonably good accuracies of cross-platform prediction approach also demonstrate its potential to solve the cold-start problem; i.e., predicting and recommending a user's activities without knowing the users' past activity history on the platform. For example, when predicting user's answer activities in Stack Overflow, we are able to achieve AUC as high as 0.71 without using any Stack Overflow features (i.e, using GitHub features \textbf{GH\_Act} only). Similar observations were made for fork, watch and favorite activities.    

We further conduct a small case study to retrieve and review fork predictions of users who did not have any past fork activities. For example, we successfully predicted that user \textit{U420338} would forked repository \textit{R12172473} in GitHub even when this was the first repository forked by the user (i.e., no past user fork activity). Examining into details, we found that \textit{R12172473} has description tags $\langle svg, javascript\rangle$, and among the 95 questions \textit{U420338} had answered in Stack Overflow, 83 contain the tags $\langle javascript \rangle$ or $\langle svg \rangle$ or both. By analyzing \textit{U420338}'s Stack Overflow activities, our approach can identify his interests, which ultimately help in predicting the user's GitHub activities. 

\textbf{Heterogeneous behaviors in cross-platform}. There are existing research work on recommendations in cross-platform setting. For example, Yan et al.\cite{yan2013friend} addressed the cold-start friend recommendation problem by combining random walk with Flickr platform features to recommend friends on Twitter. Lee and Lim \cite{lee2016friendship} performed similar cross-platform friendship recommendation in Instagram and Twitter using friendship maintenance features derived from users' friendship behaviours in multiple social platforms. In a more recent study, Lee et al \cite{lee2017analyzing} proposed a probabilistic model to predict which social media platform would a user publish a given post. 

However, most of these cross-platform recommendation studies only focus on recommending homogeneous behaviours (e.g. connecting users, publishing post, etc) in online social platforms. In our study, we handled recommendation of heterogeneous behaviors (e.g. user \textit{fork} and \textit{watch} repositories in GitHub) in a cross-platform setting. Although our prediction experiments are conducted on social collaborative platforms, the cross-platform activity prediction framework can also be applied to other online social platforms. For example, we can predict if a user will \textit{like}, \textit{comment} or \textit{retweet} a post in Twitter by learning the same user's heterogeneous behaviors in Facebook.

\section{Threats to Validity}
\label{sec:threats}
\textbf{Threats to Internal Validity}. Threats to internal validity often refer to experimenter biases. In this study, most of our processes are automated. The positive and negative instances in our training and testing datasets were also randomly selected and generated. when estimating the developer interests on GitHub, we matched the keywords in repositories' descriptions to the tags collected from Stack Overflow questions. There could be cases where the words from a repository's descriptions did not match any of the collected tags, and thus we are not able to infer the interest of the developer using that repository. In the future, we plan to mitigate this by labelling the repositories using a tag recommendation approach \cite{xia2013,stanley2013,Vargas-Baldrich2015,wang2014,xu2016Predicting}.

\textbf{Threats to External Validity}. Threats to external validity refer to the generalizability of our findings. We have used large datasets from GitHub and Stack Overflow for our analysis and experiments. Our study findings are based on 92,427 developers, which is close to 10\% of the Stack Overflow users who are active during the studied period, the interests of these users were derived using the data from the whole GitHub and Stack Overflow datasets (2.5 million and 1 million active users respectively). The current studied users were also obtained from the dataset provided in previous research by Badashian et al. \cite{badashian2014}. Both Vasilescu et al. \cite{vasilescu2013} and  Badashian et al. \cite{badashian2014} had utilized GitHub users' email address and Stack Overflow users' email MD5 hashes to find the intersection between the two platform datasets. However, this existing method is no longer valid as Stack Overflow no longer make the email hash of their user available \cite{emailhash}, thus motivating new methods to match users in two social collaborative platforms. As part of our future work, we will explore new user matching methods such as matching users by their username to generate a large dataset of cross-platform users.

\textbf{Threats to Construct Validity}. Threats to construct refer to the appropriateness of metrics used. In this work, we use precision, recall and F1-measure to evaluate the results of our prediction experiments. These metrics were commonly used for other prediction experiments, e.g.,~\cite{kim2008classifying,rahman2012recalling,nam2013transfer,canfora2013multi}.

%========================================================================================
\section{Related Work}
\label{sec:related}
In this section, we review three groups of existing research work related to our research. The first group discuss the studies on user interests in social collaborative platforms, in particularly, GitHub and Stack Overflow. The second group focuses on studies on prediction and recommendation in GitHub and Stack Overflow. The last group reviews the inter-platform studies on users in the two platforms.

The user interests in social collaborative platforms have been a widely studied research area. There were research work that focused on analyzing topics asked by the user in Stack Overflow~\cite{Barua2014,Bajaj2014,zou2015,rosen2015,yang2016security}. Barua et al. \cite{Barua2014} conducted an extensive empirical study to mine the topics discussed by users in Stack Overflow. Bajaj et al. \cite{Bajaj2014} performed a similar study specific to web development while Rosen and Shihab \cite{rosen2015} performed a similar study specific to mobile application development. Similarly, there were also work on analyzing programming languages used by users in GitHub and their relationships to GitHub contributions~\cite{Ray2014,Sheoran2014,vasilescu2015,Rahman2014}. Our work extends this group of research by comparing user interests in the two social collaborative platforms.

Prediction and recommendation in social collaborative platforms have been widely studied. These work can be further categorized into two groups: (i) finding experts to perform a certain platform tasks or activities \cite{Riahi2012,Yang2013,choetkiertikul2015,xu2016,wang2017,huang2017,yu2014,Allaho2014,yu2016,rahman2016} and (ii) recommending content or activities to users in the platforms \cite{deSouza2014,Wang2015,wang2016,guendouz2015,zhang2014,jiang2017}. For work in group (i), there were work which proposed methods to find experts to answer questions in Stack Overflow \cite{Riahi2012,Yang2013,choetkiertikul2015,xu2016,wang2017}, while for GitHub, experts are predicted if they will review \textit{pull-requests} and code for repositories \cite{yu2014,yu2016,rahman2016}. For work in group (ii), Wang et al.\cite{Wang2015} conducted a study in Stack Overflow to recommend questions and answers concerning API issues to users. de Souza et al. \cite{deSouza2014} conducted an experiment to recommend Stack Overflow question-answer pairs relevant to selected software programming problems. Zhang et al. \cite{zhang2014} predict and recommend relevant repositories to users based on the users' past activities (e.g. fork, watch, etc) in the platform. In a more recent work, Jiang et al. \cite{jiang2017} proposed to use user programming language preferences and one-class collaborative filtering to improve prediction of which GitHub repositories are relevant to a user. Our study adds on to the state-of-the-art in group (ii) by proposing a novel method that use {\em multiple} platform data to predict platform activities

There have been few existing inter-platform studies on GitHub and Stack Overflow. Vasilescu et al. performed a study on developers' involvement and productivity in Stack Overflow and GitHub~\cite{vasilescu2013}. They found that users who are more active on GitHub (in terms of GitHub commits), tend to ask and answer more questions on Stack Overflow. Badashian et al. \cite{badashian2014} did an empirical study on the correlation between different types of activities in the two platforms. Silvestri et at. \cite{silvestri2015} proposed a user linkage model to link users' Stack Overflow, GitHub and Twitter accounts. More recently, Lee and Lo \cite{lee2017} did an extensive study on users' interests across GitHub and Stack Overflow. In that study, the researchers found that users who have accounts with GitHub and Stack Overflow do exhibit similar interests observed from their activities in the two social collaborative platforms. Our work builds upon insights reported in \cite{lee2017} by proposing a prediction method based on user interests across and within social collaborative platforms.  

%========================================================================================
\section{Conclusion and Future Work}

\label{sec:conclusion}
In this paper, we propose a novel framework which predicts users activities in multiple social collaborative platforms. We conducted experiments on large real-world datasets which contain activities of 92,427 users who are active in GitHub and Stack Overflow. Our proposed methods achieved good accuracy in predicting various user activities (up to an AUC score of 0.89).

Our experiments have shown that user activities in Stack Overflow can be predicted with reasonable accuracy using the same user's interests inferred from his or her activities in GitHub. The same observation was made when predicting a user's activities in GitHub using his or her interests inferred from his or her activities in Stack Overflow. The reasonable accuracies yield by cross-platform prediction approach demonstrates its potential in solving the cold-start problem in user activity prediction and recommendation in social collaborative platforms.

For future work, we intend to consider more advanced techniques (e.g., topic models or deep learning models) to derive and measure user interests similarity across multiple platforms. We will also consider platforms aside from Stack Overflow and GitHub (e.g. Quora).

\section*{Acknowledgments}
This research is supported by the Living Analytics Research Centre (LARC), a research centre set up in Singapore Management University (SMU) with Carnegie Mellon University (CMU) that focuses on behavioural and social network analytics in the fields of Urban and Community Liveability, Personalised Urban Mobility, Jobs and Skills Intelligence, as well as Smart Consumption and Healthy Lifestyle.

\balance
\bibliographystyle{ACM-Reference-Format}
\bibliography{ref} 

\end{document}